\documentclass[letters,fleqn,usenatbib]{mnras}

\usepackage{newtxtext,newtxmath}

\usepackage[T1]{fontenc}
\usepackage{ae,aecompl}

\usepackage{graphicx}	
\usepackage{amsmath}	
\usepackage{amssymb}	

\title[The Metallicity-Period-Mass Diagram of low-mass exoplanets]{The Metallicity-Period-Mass Diagram of low-mass exoplanets}

\author[S. G. Sousa et al.]{
S\'ergio G. Sousa,$^{1}$\thanks{E-mail: sergio.sousa@astro.up.pt (SGS)}
Vardan Adibekyan,$^{1}$     
Nuno C. Santos,$^{1,2}$
Annelies Mortier,$^{5}$             
\newauthor
Susana C. C. Barros,$^{1}$          
Elisa Delgado-Mena,$^{1}$
Olivier Demangeon,$^{1}$
Garik Israelian,$^{3}$
\newauthor
Jo\~ao P. Faria,$^{1}$
Pedro Figueira,$^{4,1}$
Barbara Rojas-Ayala,$^{6}$
Maria Tsantaki,$^{1}$              
\newauthor
Daniel T. Andreasen,$^{1}$
Isa Brand\~ao,$^{1}$
Andressa C. S. Ferreira,$^{1}$
\newauthor
Marco Montalto,$^{7}$
Alexandre Santerne,$^{8,1}$
\\
$^{1}$Instituto de Astrof\'isica e Ci\^encias do Espa\c{c}o, Universidade do Porto, CAUP, Rua das Estrelas, 4150-762 Porto, Portugal\\
$^{2}$Departamento de F\'isica e Astronomia, Faculdade de Ci\^encias, Universidade do Porto, Rua do Campo Alegre, 4169-007 Porto, Portugal\\
$^{3}$Instituto de Astrof\'isica de Canarias, 38200 La Laguna, Tenerife, Spain\\
$^{4}$European Southern Observatory, Alonso de Cordova 3107, Vitacura, Santiago, Chile\\
$^{5}$Astrophysics Group, Cavendish Laboratory, J.J. Thomson Avenue, Cambridge CB3 0HE, UK\\
$^{6}$Departamento de Ciencias Fisicas, Universidad Andres Bello, Fernandez Concha 700, Las Condes, Santiago, Chile\\
$^{7}$Dipartimento di Fisica e Astronomia "Galileo Galilei", Universit\'a di Padova, Vicolo dell'Osservatorio 3, Padova IT-35122, Italy\\
$^{8}$Aix Marseille Univ, CNRS, CNES, LAM, Marseille, France
}

\date{Accepted XXX. Received YYY; in original form ZZZ}

\pubyear{2018}

\begin{document}
\label{firstpage}
\pagerange{\pageref{firstpage}--\pageref{lastpage}}
\maketitle

\begin{abstract}
   The number of exoplanet detections continues to grow following the development 
   of better instruments and missions. Key steps for the understanding of 
   these worlds comes from their characterization and its statistical studies.
   We explore the metallicity-period-mass diagram for known exoplanets by using an updated version of The Stellar parameters for stars With ExoplanETs CATalog (SWEET-Cat), a unique compilation of precise stellar parameters for planet-host stars provided for the exoplanet community.
   Here we focus on the planets with minimum mass below 30 M$_{\oplus}$ which seems to present a possible correlation in the metallicity-period-mass diagram where the mass of the planet increases with both metallicity and period. Our analysis suggests that the general observed correlation may be not fully explained by observational biases. Additional precise data will be fundamental to confirm or deny this possible correlation.
\end{abstract}




\section{Introduction}

\begin{figure*}
  \centering
  \includegraphics[width=19cm]{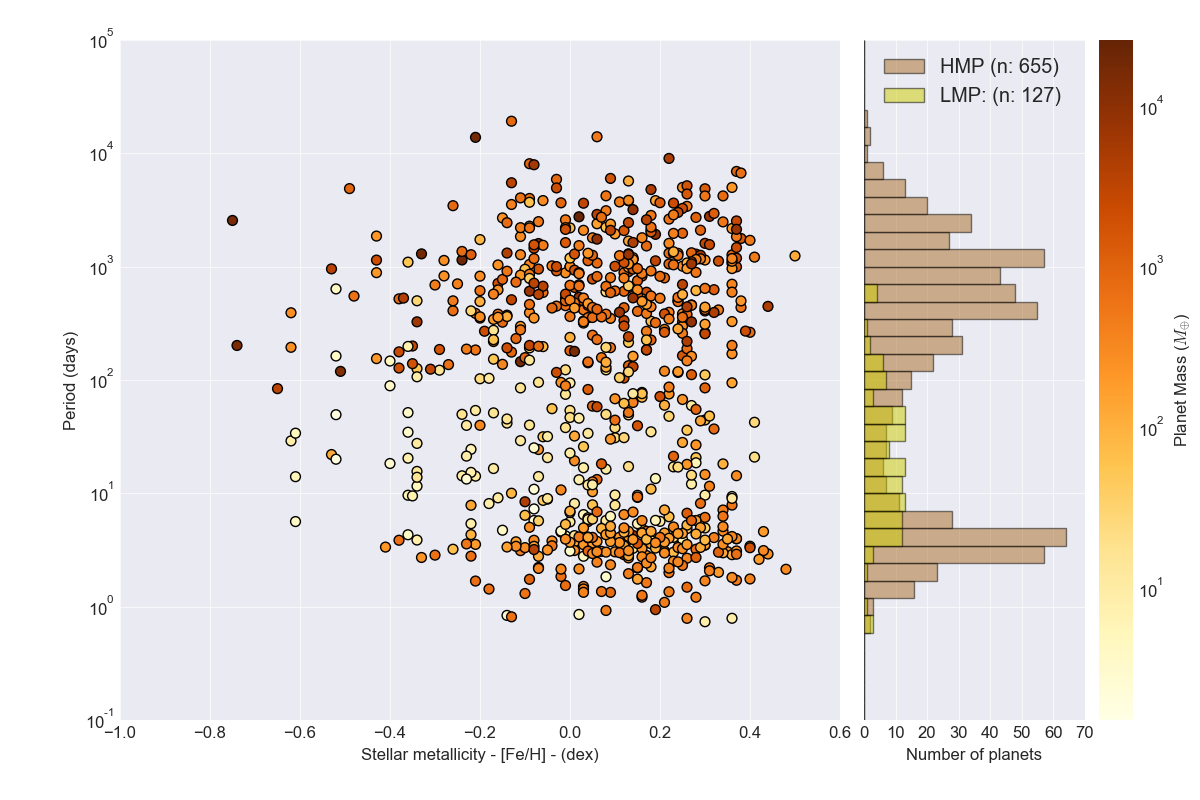}
  \caption{The Period-Metallicity diagram for planets with homogeneous parameters in SWEET-Cat. The color scheme represents the planet mass (in $M_{\oplus}$). On the right panel we plot the period distribution for HMP (brown) and LMP (yellow).}
  \label{the_figure}
\end{figure*}

One of the first observational constraints for planet formation theories was the correlation between the presence of giant planets and the metallicity of their host stars where the massive planets are more common around metal-richer stars \citep[e.g.][]{Santos-2004b, Fischer_Valenti-2005}. Given the known observational biases of the most successful detection methods (radial velocities and transits), that are sensible to either massive and larger planets at short periods, the first planets to be discovered were as or more massive than Jupiter with periods of only a few days. As the techniques improved with the time passed, exoplanets with masses below that of Jupiter were found, and the planet metallicity correlation was tested for neptune-like and rocky exoplanets, where, despite the very low statistical significance of the first results, the metallicity correlation seemed to be absent and not following what has been observed for their higher mass counterparts \citep[e.g.][]{Udry-2006, Sousa-2008, Ghezzi-2010, Sousa-2011, Buchhave-2012}. Recently, \citet[][]{Wang-2015} suggested that there should be a universal planet-metallicity correlation for all planets, however, this result might be related to the higher planet frequency and to the lower detectability of low-mass planets \citep[][]{Zhu-2016}. Very recently \citet[][]{Petigura-2018} sugested a positive correlation of the planet occurrence rates with metallicities including small planets, except for the very small, Earth-size, planets for which the correlation seems not to be present. These trends are compitible with the works of \citet[][]{Buchhave-2014, Buchhave-2015}. All these observational correlations provide strong constraints for the theory of planet formation and evolution.

In this work we will take advantage of a large sample of planet-host stars with homogeneous spectroscopic parameters of the Stars With ExoplanETs CATalog (see SWEET-Cat; \citet[][]{Santos-2013}). We will use the latest version of this catalogue which was recently updated in \citet[][]{Sousa-2018} and in combination with the planet properties listed in exoplanet.eu \citep[][]{Schneider-2011} we study in detail the metallicity-period-mass diagram of exoplanets in search of observable correlations.

This paper is divided into the following sections: In section 2 we present the metallicity-period-mass diagram. In Section 2.1 we focus the discussion on the low-mass planets and report an interdependence of the parameters. In section 2.2 it is checked if multiple planetary systems may explain the result. Then in Section 2.3 we re-check the possible presence of an upper boundary in the mass-metallicity planet for low mass planets. In Section 3 we present a discussion mentioning any possible observational bias, and then considering some basic ideas for planet formation and evolution. Finally in Section 4 we summarize this work.

\section{Metallicity-period-mass diagram}

The metallicity of the host star and the period and mass of the exoplanet are explored for several exoplanets in the metallicity-period-mass diagram (MPM diagram) presented in Figure \ref{the_figure}. The SWEET-Cat catalog is characterized for the homogeneity of the spectroscopic stellar parameters found in it, and we only considered planet-hosts with Homogeneity flag $=$ 1. This means that the spectroscopic stellar parameters were derived using the same methodology (for more details see \citet[][]{Sousa-2014}). To avoid finding trends that may be related to the precision on the determination of spectroscopic stellar parameters, instead of related to the physics of planet formation and evolution, we also did a cut in effective temperature. We thus only included planets for which the host star have Teff $>$ 4500K in the MPM diagram, given that the spectroscopic stellar parameters are less precise for the cooler stars ($\sigma >= 0.1$ dex in [Fe/H], e.g. \citet[][]{Rojas-Ayala-2012, Mann-2013, Onehag-2012}). 
A total number of 782 exoplanets are included in the MPM diagram. In the right panel we also show the period distributions of the exoplanets, where we divided the sample in two by exoplanet mass, labelling as High-Mass Planets (HMPs - 655 exoplanets) the ones with minimum masses greater than 30 M$_{\oplus}$ ( $\sim 0.095$ M$_J$), and Low-Mass Planets (LMPs - 127 exoplanets) the ones with minimum masses less or equal to 30 M$_{\oplus}$. The value
of 30 M$_{\oplus}$ comes from the location of the gap in the planet mass distribution presented in \citet{Mayor-2011} and we keep this value to separate in mass the samples to be consistent with our previous
works \citep[e.g.,][]{Sousa-2011b}. Note that we have also considered a separation at 20 and 40 M$_{\oplus}$) however no significant changes were seen with these different values to cut into two different distributions.

The division of the sample in two groups by exoplanet mass reveals three populations of exoplanets in the distribution of periods. The HMPs distribution shows 2 peaks at different periods: one corresponds to the very well known ``hot Jupiters'' population with the peak in period around 3 days \citep[e.g.][]{Dawson-2018}, and the second corresponds to Jupiter-like planets at longer periods. This last one presents a wider dispersion of periods with a peak close to 1000 days. 

The short period peak followed by a ''valley`` in the period distribution of giant planets has already been identified and discussed in several works \citep[e.g.][]{Dawson-2018}. Observational bias could be responsable for the existence of the ''valley`` given that the hot Jupiters are mainly discovered by transists, while the jupiter-like planets are detected with RVs. This result was already discussed as based on older radial-velocity surveys \citep[][]{Udry-2007b}. Despite the strong bias of the transit technique to detect shorter period planets, recent results have also confirmed that such trend exists in Kepler data. For example, Fig. 8 of \citet[][]{Santerne-2016b} shows a very similar picture stating that although this peak of hot-Jupiters was not confirmed in previous analysis of Kepler data, it reveals itself once the false positives are removed from the Kepler data.

The LMPs distribution is located in between these two populations of HMPs. We will focus our discussion in the distribution of LMPs from now on.

\subsection{The low-mass exoplanet distribution}

The observed distribution of the LMPs in Figure \ref{the_figure} seems to be different from the HMPs: the LMPs preferentially occupy the gap of space between the two peaks of the HMPs distribution. Due to detectability limits, we are not able to have a complete picture of the period distribution of LMPs at longer periods, but it is evident that this population will be different anyway, given that the HMPs with periods of 10-100 days can be detected easily by RV surveys, but they are under represented in the HMPs distribution. Also the fact that low mass planets have a different distribution (for short periods) than giant planets has already been discussed in other works \citep[e.g.][]{Boue-2012}.

\begin{figure}
  \centering
   \includegraphics[width=9.5cm]{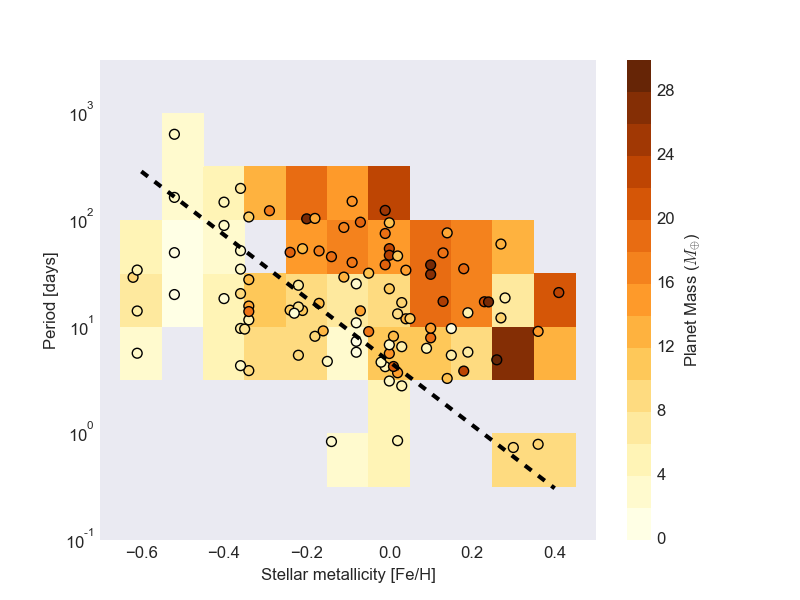} \\
  \caption{The same diagram as in Fig. \ref{the_figure} but only for the low-mass planets ($ < 30 M_{\oplus}$) with planetary masses derived with at least 20\% precision. The color scheme represents the planet mass (in M$_{\oplus}$). The colored background corresponds to the diagram binned average planet mass. The dashed black line is the linear fit of the data for a constant mass of 10M$_{\oplus}$.}
  \label{the_figure_low_mass}
\end{figure}

In this work we will focus on the detected LMPs whose period distribution appears to be quite disperse, without a clear maximum in its distribution as seen in Figure \ref{the_figure}. Note however that we are not claiming here that the real LMP period distribution is flat. To recover the real (biases corrected) period distribution of low mass planets one need to consider very carefully all the detection limits coming from the different detection surveys, which is out of the scope of the current study. The reason for the apparent (visual) flat distribution might be due to the small sample size of the LMPs and at some level to the relative scale of the figure where the LMPs and HMPs period distributions are shown together.

Note however that when we remove the constraint for LMPs listed in SWEET-Cat with Homogeneity flag $=$ 1 the period distribution appears different with a clear peak around 10 days. This peak at 10 days is most likely linked with transit observation bias which favour the detection of short period planets. Note also that many of these LMPs were detected by transit surveys (e.g. Kepler mission) which target many faint stars. Because of this SWEET-Cat does not have an homogeneity flag of 1 (i.e., they were excluded from our analysis). The LMPs period distribution presented in figure \ref{the_figure}, where 80 \% of the planets were detected by RV surveys, is actually very close to the one presented in Figure 14 of \citet[][]{Mayor-2011} when considering only RV detections for a specific survey. In the same figure of \citet[][]{Mayor-2011} we can see how the real period distribution of LMPs is affected when considering detection limits.

Focusing on the LMPs with homogeneously derived stellar metallicities, we present a zoomed-in version of the MPM diagram where only reliable LMPs are included. Therefore, in Figure \ref{the_figure_low_mass} we only include planets with planetary masses derived with an absolute precision better than 6 M$_{\oplus}$, i.e. below 20\% accuracy for a planet at the threshould of 30 M$_{\oplus}$. This absolute precision constraint allows to keep 103 LMPs. We choose this approach because when constraining the sample with a relative precision (e.g. 20\%) we were removing several very low-mass planets with slightly higher uncertainties but which are for sure LMPs with masses below 30 M$_{\oplus}$. Figure \ref{data_errors} presents the 3-dimensional errors for each LMP in the sample.

A gradient in mass is observable going from the lower mass LMPs (yellow points) in the metal-poor regime (left side of the diagram) to the most massive LMPs (orange/red points) on the metal-rich regime (right side of the diagram). To make it more visible in Figure \ref{the_figure_low_mass} the background color represents a binned averaged planetary mass which is computed using an implementation of a boxcar average in 2D. This representation shows a planet mass dependence not only on metallicity but also on the period of the planets. The masses of the LMP sample seems to grow with the increase of the metallicity of the host star and with the increase of their orbiting period. To quantify the level of dependence of the visual correlation in the data, we performed a 3D plane fit to the data considering the errors for all the variables (see Appendix A for more details about the fitting method and respective references). The coefficients that we find when fitting this data are the following:

\begin{equation}
 M_{p} = (5.3 \pm 0.2) + \\
              (12.0 \pm 0.5) * [Fe/H] + \\
              ( 4.0 \pm 0.2) * \log(P)
\label{eq_mass}
\end{equation}

where Mp is the mass of the planet, [Fe/H] is the metallicity (in dex) and P the period of the planet (in days). In Figure \ref{the_figure_low_mass} we also show this fit when fixing the mass to 10 M$_{\oplus}$ for a better visual representation of the correlation. In Section 3 we discuss in detail possible interpretations of this correlation and possible observational biases which can be in play.

\subsection{Multiple planetary systems with low-mass exoplanets}

The effect of planet-planet interactions can definitely affect the evolution of the formed planets in multi-planetary systems. Massive Jovian planets specially certainly have major impact on the dynamics of the small planets in the systems \citep[e.g.][]{Kobayashi-2012, Becker-2017, Munoz-Romero-2018}.
A significant part of the analysed LMPs are indeed in multi-planetary systems (only 22 out of 127 LMPs and 16 out of 84 LMPs, are currently single in Figure \ref{the_figure} and \ref{the_figure_low_mass}, respectively). For the metal-rich regime, it is reasonable to consider that the presence of massive planets in the system, preferentially found in short (3-day) and long (1000 day) periods, may have a strong impact on the dynamics of the systems, affecting the orbital evolution and the observed periods of the LMPs in multi-planetary systems. This could introduce an effect on the MPM correlation that we described in the previous sections.

To check for this sort of bias, we looked at two subsamples of the LMPs. The first subsample is composed of LMPs which are in single-planet systems or for which all the planets in the system have masses below 30 $M_{\oplus}$. The second subsample are the rest of the LMPs which have at least one massive planet as companion on their multi-planetary systems. If planet-planet interaction plays a role in the observed correlation, the two subsamples are expected to show significant differences in the MPM diagram. Interestingly, this analysis has shown that there are no significant changes in the trend in the MPM diagram after the removal of the LMPs with massive companions, and the least-square fitting returns similar coefficients when compared with the ones in equation \ref{eq_mass}. 

\begin{figure}
  \centering
   \includegraphics[width=9.0cm]{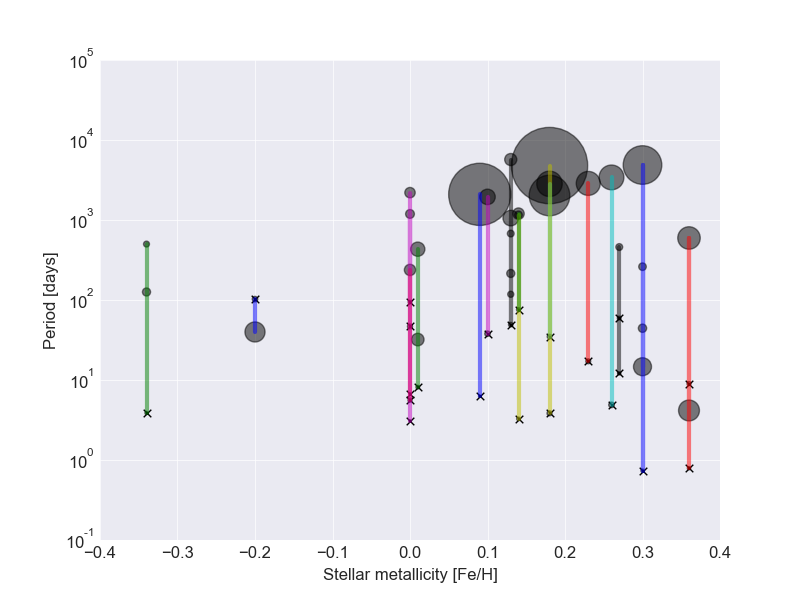} \\
  \caption{Multi-planetary systems that contain at least one low mass planet and at least one giant planet. The crosses represent the low mass planets, while the semi-transparent circles represent the higher mass planets where the size of the circles scales with the planetary mass.}
  \label{sistems_homog}
\end{figure}

For the second sample, which contains only the LMPs with massive planets in their multi-planet system, we use a slightly different representation. There are only 22 LMPs (26\% of all LMPs) in the second sample and they are shown in the metallicity-period diagram of Figure \ref{sistems_homog}. The planets of the same system are connected with distinct vertical lines for clarity. The LMPs are represented with `x' while the giant/massive planets are represented with a semi-transparent circle, which size is proportional to their planetary mass. Given the metallicity correlation of massive exoplanets, these systems are preferentially found around metal-rich stars, and, hence, we do not have many systems at low metallicity in our second sample. Although the number of these systems is small, these systems have preferentially their LMPs in shorter periods when compared with the massive planet companions. This is specially evident at higher metallicities. In a different perspective, our sample shows that Hot Jupiters are unlikely to have LMPs detected in relatively short periods (at least where the LMPs can be detected). In fact, there is only one Hot Jupiter (P < 10 days) at the right part of Figure \ref{sistems_homog} (WASP-47b). This is consistent with previous results in what regards Hot Jupiters, since it is known that they are mostly alone on their planetary systems \citep[][]{Steffen-2012, Huang-2016, Schlaufman-2016}. 

Since the two sub-samples do not show significant changes in the MPM diagram we can deduce then that the dynamics of multi-planetary systems do not have a measurable effect on the mass-metallicity-period dependence. The discovery of new low-mass planets with precise characterization will certainly help in confirming this observational result.

\subsection{An upper boundary in the mass-metallicity exo-Neptunes}

\begin{figure}
  \centering
  \includegraphics[width=9.0cm]{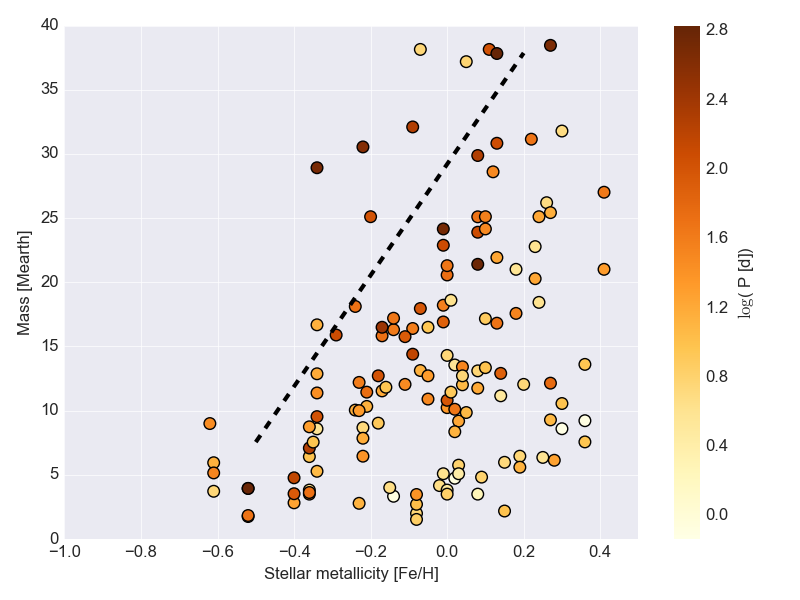}
  \caption{Planetary mass vs. the metallicity. This is a reproduction of the top panel of Fig. 3 of \citet[][]{Courcol-2016} with updated SWEET-Cat data. Dashed line 
  corresponds to the proposed upper boundary on the same work. The color scale represents the period (days) of the planets in logarithm scale.}
  \label{mass-metallicity}
\end{figure}

A possible upper boundary in the mass-metallicity of exo-Neptunes was proposed by \citet[][]{Courcol-2016}. This upper boundary, if real, could give important constraints for the formation of LMPs assuming that the metallicity is a key parameter for the maximal mass of these small planets. Although, the work by \citet[][]{Courcol-2016} was clearly focused on the maximum mass of the planet and not on the general distribution of the mass in the mass-metallicity diagram, it is also mentioned a dependence with the orbital period which is consistent with our MPM relation presented before.

Using the SWEET-CAT data we reproduce in Figure \ref{mass-metallicity} an updated diagram of that shown in \citet[][Figure 3]{Courcol-2016}. Note that in \citet[][]{Courcol-2016} low-mass planets were consider up to 40 M$_{\oplus}$. The proposed upper limit is shown as a dashed black line. Most of the added planets in the updated diagram are indeed below the dashed line, but there are few new planets above the proposed upper boundary. Most of these ''outliers`` are planets with masses above 30 M$_{\oplus}$. This lead us to question the mass values to select the exo-Neptunes and that identify this upper boundary. In \citet[][]{Courcol-2016} the cut was performed at 40$_{\oplus}$, and at the time of that work, there were just two planets found just above the upper boundary line. When we consider more massive planets, several points appear above this line, but then we are entering into a different regime of planets (sub-Jovians). Therefore, the definition of such an upper boundary, if real, will strongly depend on the selection of the planets and this is clearly not an easy task given the small number of low-mass planets. More data, or at least, more precise data, is needed to confirm, reject, or better define a possible upper limit boundary.

\section{Discussion}

\subsection{Observational bias}

Observational bias of the detection methods to find exoplanets could play an important role in the interdependence of the parameters that we observe here in the MPM diagram, specially when focusing on low-mass planets. The vast majority of these LMPs are detected with the radial-velocity technique (99/127 in Fig. \ref{the_figure} and 60/84 in Fig. \ref{the_figure_low_mass}), where the higher the mass of the planet, the easier it is to detect it. This could partially explain the dependence on the mass with the period that we is present in the diagram.

\begin{figure}
  \centering
  \includegraphics[width=9cm]{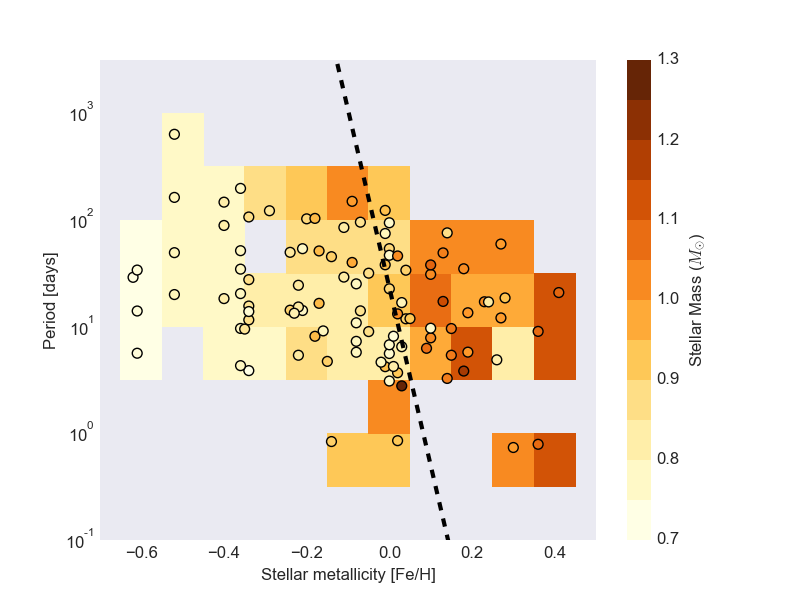} 
  \includegraphics[width=9cm]{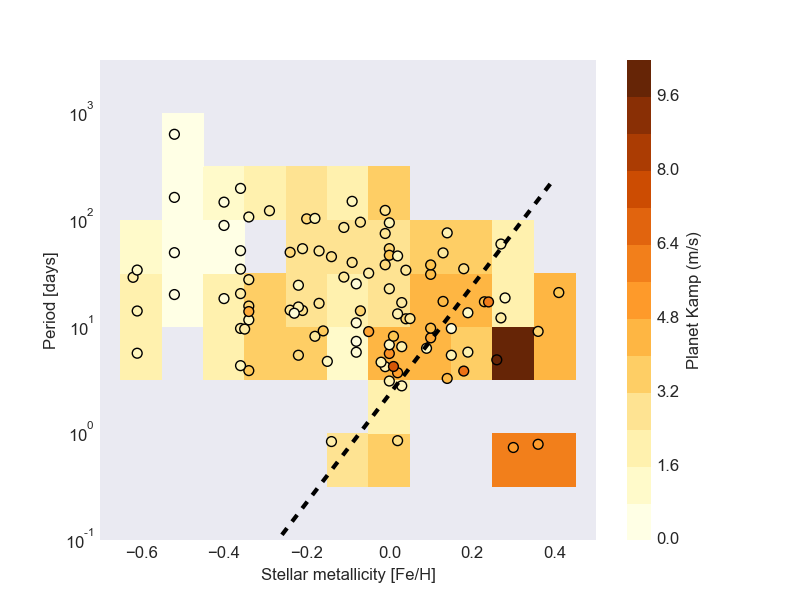}
  \caption{Same as Figure \ref{the_figure_low_mass} but the color scale represent the stellar mass (top panel) and the RV semi-amplitude (bottom panel). The dashed line represents the linear fit for a constant value, 0.9 M$_\odot$ and 4 m/s on the top and bottom panels respectively.}
  \label{fig_bias}
\end{figure}

On the metallicity dependence in the diagram, we know that the metallicity has a small impact on the radial-velocity detections \citep[see e.g. Figure 8 of ][]{Valenti-2008}. However when looking carefully at our sample there is another relevant dependence with metallicity which has a direct impact on the RV detection technique. The stars hosting LMPs have a stellar-mass vs. metallicity correlation, where the most metallic hosts are more massive than the metal-poor hosts. In the top panel of Fig. \ref{fig_bias} we plot the same kind of MPM diagram, but exchanging the planet mass by the host stellar mass. The background color is the box car average of the stellar mass, and the dashed line marks a linear fix for a constant stellar mass of 0.9 $M_{\odot}$. This dependence will also affect the detection of the lower mass planets, specially for the metal-rich stars in our sample. At this point there are two interesting questions that can be raised. The first is: Are these LMPs host stars different from typical stars? And, can this correlation be enough to explain the interdependence that we see in the data for the MPM diagram?

To answer the first question, we compared our sample of LMPs host stars against a sample of 1111 stars located in the solar neighborhood taken from the HARPS sample discussed in \citet[][]{Adibekyan-2012b}. In fact in \citet[][]{Sousa-2018} we already compared this solar neighborhood sample against the full sample of LMPs host stars showing that in what regards the metallicity distribution they are in fact indistinguishable. Now, when comparing the distribution of the stellar mass in the samples we also do not see any significant differences, specially given the small number of LMPs host stars. Performing a Kolmogorov-Smirnov (K-S) test on both samples we obtain K-S statistic of 0.15 and a p-value of 7.95e-02. And if we select only the metal-rich stars from these samples ([Fe/H] > 0), i.e. for the MPM region where we are ''missing`` the less massive LMPs), we see a K-S statistic of 0.16 and a p-value of 1.66e-01. Therefore we can conclude that as we see for metallicity, here we also don't see any relevant difference between the stars hosting LMPs and the neighborhood stars in what regards the stellar mass. It is still reasonable to point out the average and standard deviation of the masses that we have at low and high metallicities. For metal-poor stars ([Fe/H] < 0) from the neighborhood sample we see the average stellar mass of  0.87 $\pm$ 0.17 M$_\odot$ while for the metal-rich stars we have an average stellar mass of 1.02 $\pm$ 0.18 M$_\odot$. Although the average is significantly higher, the standard deviation shows that we can still have stars with relatively low mass at higher metallicities, and for which we could detect lower mass planets with the current instrumental limitations. 

This last statement partially answers the second question. However to be more focus on our sample of LMPs, we choose to plot on the bottom panel of Figure \ref{fig_bias} the RV semi-amplitude for all the LMPs in our sample. The RV semi-amplitude was estimated in the same way for all the LMPs since it was not reported in exo.eu for a non negligible number of exoplanets. These values were then estimated using the stellar masses, the minimum planetary masses, the orbital period and the the eccentrities (where we assumed zero when it was not available in exo.eu). As for MPM diagram in figure \ref{the_figure_low_mass} we also use a box car average for the background color for a better visualization. A linear fit was also performed to this figure, where the dashed line mark the line where we have a RV semi-amplitude of 4 m/s. From the figure it is clear that for the planets around metal-rich stars, these have on average higher semi-amplitude when compared with the metal-poor host stars, meaning that in fact we would be able to detect less massive planets in this region of the MPM. This means that although there is a clear observational bias, both for period, and metallicity, it is probably not enough to explain the lack of very-low mass and detectable planets in the metal-rich region of the MPMs. Therefore, it is difficult to explain a mass-period-metallicity dependence only by these observational bias.

One final note on observational bias that we would like to mention is related with stellar activity. Given that the RV semi-amplitudes are larger for metal-rich stars we also consider the possibility that the stellar activity, for some reason, would be greater for the metal-rich stars, and could pull the RV limits to lower precision. However, to check this carefully we would need to go star-by-star to check individual stellar activity. Moreover, this issue should be very unlikely to be a problem. From one side, the RV planet-search samples are already very conservative in what regards stellar activity, and we remind the reader that the large fraction of these planets were detected by RV. On the other side, the stellar activity is not always a problem for the planet detection, since the planetary signals are in many of the cases completely disentangle from the stellar activity signal, specially for the cases where the stellar rotation period is significantly different from the planet orbital period.

\subsection{Correlation by chance?}

So far we have discussed several observational biases that can affect the correlation that we present in this work. In the last section we refer to a few observational biases which can affect the correlation presented in Fig. \ref{the_figure_low_mass}. In this section we try to quantify the effect of these  observational biases, or in other words, estimate the probability that the observed correlation is obtained by chance.

To address this question we performed Monte-Carlo simulations to statistically represent the MPM diagram based on the following assumptions:

\begin{itemize}
 \item The simulated hosting star (stellar mass, and stellar [Fe/H]) is randomly selected from the HARPS sample discussed in \citet[][]{Adibekyan-2012b}. For each star we assume a Gaussian error for its mass and metallicity (10\% in mass and 0.05 dex in metallicity). This way we are keeping the stellar mass and stellar metallicity distributions that we have for the solar neighborhood including the correlation between these two variables. Note again that the metallicity distribution of LMPs is basically indistinguishable from the solar neighborhood stars metallicity distribution;
 
 \item The planet minimum mass is selected from a uniform distribution between 0.25 and 30 M$_{\oplus}$;
 
 \item The planet logarithmic period is selected from a uniform distribution between 0 and 2.2 (1 day and $\sim$ 160 days - the range of periods observed in Figure \ref{the_figure_low_mass}). With this we are ensuring that the distributions of the points in the MPM diagram will be very similar to the one that we have for the real data;
 
 \item For the estimation of the RV semi-amplitude we assume planet with eccentricity 0.
\end{itemize}
We then did 1000 random draws from these distributions and for each resulting star-planet system we computed the expected semi-amplitude of the radial velocity signal. Only planets producing an RV semi-amplitude above 1 m/s were selected. This is the typical RV precision that we have in many RV surveys (e.g. using HARPS). Each simulated sample had 100 points in order to be consistent to the size of the real data.

\begin{figure}
  \centering
  \includegraphics[width=8cm]{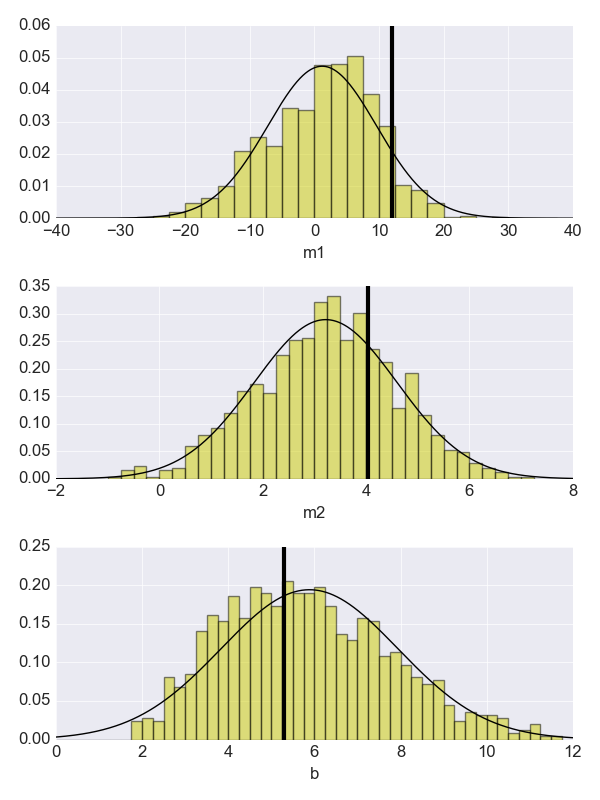} 
  \caption{3D planet coefficient distributions derived from the simulations. A gaussian fit is presented for each distribution. The vertical black line marks the slope derived for the real data.}
  \label{simulated_slopes}
\end{figure}

These assumptions allow to generate data considering the most important observational bias discussed in the previous section. For each randomly generated sample we then used the same 3D fitting method to extract the correlation slopes (m1 - slope for [Fe/H] in units [M$_{\oplus}$/dex], and m2 - slope for log(period) in units [M$_{\oplus}$/dex]) for each simulation and compare the slopes distributions to the slopes derived for the real data. The goal is to try to estimate the probability that the slopes obtained for the real data could come by chance. An example of the MPM diagram for one of these simulations is presented in the top panel of Figure \ref{fig_simulated}.


In figure \ref{simulated_slopes} we present the distributions of the slopes m1 and m2 obtained for our simulations. The average slope m1 is 1.20 $\pm$ 8.43 [M$_{\oplus}$/dex] and the average slope m2 is 3.21 $\pm$ 1.38 [M$_{\oplus}$/dex]. From the respective distributions we can state that the metallicity correlation (m1) due to the observational bias seems to be very small, but with high uncertainty. The slope that we obtain for the real data (12.0 [M$_{\oplus}$/dex]) is outside the one sigma standard deviation, making it unlikely to retrieve this value by chance under our assumptions. In what regards the period correlation (m2), the expected correlation due to observational bias seems to be quite high and more precise when compared with m1. The value (4.0 [M$_{\oplus}$/dex]) that we get for the real data is within one sigma. Therefore we cannot exclude the possibility that the period correlation found in the real data could indeed be entirely due to the assumed observational bias.

A similar exercise was also done to check the 3D fits when using the RV semi-amplitudes instead of the planetary masses for the simulated data (see bottom panel of figure \ref{fig_simulated}). From the simulations these slopes distributions show an average slope m1 of -2.19 $\pm$ 2.59 [m/s/dex], and an average slope m2 of -3.82 $\pm$ 0.59 [m/s/dex]. These slopes better represent and quantify the expected observational biases due to RV detection limits. Interestingly for the real data presented in the bottom panel of figure \ref{fig_bias} we obtained m1 = 3.62 $\pm$ 0.45 [m/s/dex] and m2 = -0.71 $\pm$ 0.19 [m/s/dex]. Both these slopes are  unlikely to be derived by chance.

With these simulations we show that it is unlikely to find the presented general 3D correlation by chance. However some caution should be taken when considering the period correlation. Additional precise data will be fundamental to confirm or infirm this possible interesting correlation.

\subsection{Clues for planet formation and evolution}

There is plenty to learn in what regards the formation and evolution of planets. Several theoretical mechanisms in the literature can be used to explain the observations. How does our observed mass-period-metallicity dependence on LMPs fit in some of the most trending theoretical ideas?

The migration of planets is a theoretical mechanism quite discussed in the literature that should act more efficiently on massive planets, given that the mass of the planet have a direct impact on the size of the gap in the circumstellar disk, affecting directly the time scale of migration \citep[][]{Lubow-2010, Baruteau-2016}. If we consider migration for the LMPs, this could be somehow compatible with the correlation that we observe in the MPM diagram. These LMPs are disperse in period, but their mass and metallicity of the system seem to play a role in the final position of their orbits. We could consider that the planet mass can slow down the effect of migration. But if migration is the only mechanism in play, the observed trend tells us that metallicity would also affect the migration. The higher the metallicity the longer should be the time scales for migration, allowing for the more massive LMPs to be closer to the star for higher metallicity environments.

In-situ formation for LMPs \citep[e.g.][]{Chiang-2013} can be another mechanism to explain the dependence that we present here. Simply considering the assumption of an homogeneous formation disc, we would expect the observed mass-metallicity-period correlation. The increase of metallicity would mean higher presence of the fundamental building blocks to form the planets. The closer the formation site (radius/period) is to the star, the less quantity of condensed material closer to the star will be available to form the planet. Therefore, for shorter periods, we would have lower mass planets, just like we observe in Figure \ref{the_figure_low_mass}.

A combination of these two effect could also be at play. One certain remains, that with the still low number of low-mass planets with precise masses detected and the several observational bias that can be in place, it is clear that we need more detection of LMPs to better understand and quantify this MPM interdependence.

%
\section{Summary}

We use the recently updated version of SWEET-Cat to exploit the metallicity-period-mass (MPM) diagram of known exoplanets. In the diagram the detected exoplanets are clustered in three groups: the well known hot-Jupiter population; a population of jupiter-like massive planets at long periods; and a low mass planet population. We focus our work on the detected low-mass planet population that have periods in between the previous two massive planet populations.

We found a dependence of the mass of low-mass planets on the metallicity and the period, which is close to linear, correlating both with the metallicity and the logarithm of the period. The mass of the planets increases for metal richer stars and for longer periods. 

Although the dynamics of multi-planetary systems influences the architectures of planetary systems, a very simple test was performed looking at two subsamples of planetary systems (only low mass planet systems vs low mass planet with massive companion) revealing no significant differences in the observed MPM dependence.

We also review the proposed upper boundary in the mass-metallicity plane of exo-Neptunes proposed by \citet[][]{Courcol-2016}. We show that there are some new low-mass planets above the proposed line, but more data is required to better define the possibility of this upper boundary.

The possible correlation observed in the MPM diagram, can be strongly affected by observational biases, but we show that these biases alone cannot entirely explain the general interdependence observed in the MPM diagram. More precise data will be fundamental to confirm this possible interesting correlation which can provide important constraints to the theories of planet formation and evolution.

\section*{Acknowledgements}

This work was financed by FEDER - Fundo Europeu de Desenvolvimento Regional funds through the COMPETE 2020 - Operacional Programme for Competitiveness and Internationalisation (POCI), and by Portuguese funds through FCT - Funda\c{c}\~ao para a Ci\^encia e a Tecnologia in the framework of 
the projects POCI-01-0145-FEDER-028953, PTDC/FIS-AST/7073/2014 \& POCI-01-0145-FEDER-016880, and  PTDC/FIS-AST/1526/2014 \& POCI-01-0145-FEDER-016886. P.F., S.C.C.B., N.C.S., S.G.S, V.A. and E.D.M. acknowledge support from FCT through Investigador FCT contracts nr. IF/01037/2013CP1191/CT0001, IF/01312/2014/CP1215/CT0004, IF/00169/2012/CP0150/CT0002, IF/00028/2014/CP1215/CT0002, IF/00650/2015/CP1273/CT0001, and IF/00849/2015/CP1273/CT0003. S.G.S. further acknowledges support from FCT in the form of an exploratory project of reference IF/00028/2014/CP1215/CT0002. ACSF is supported by grant 234989/2014-9 from CNPq (Brazil). J.P.F. and acknowledges support from FCT through the grants SFRH/BD/93848/2013 and SFRH/BPD/87857/2012. B.R.-A. acknowledges the support from CONICYT PAI/Concurso Nacional Inserci\'on en la Academia, Convocatoria 2015 79150050. O.D. also acknowledges support from FCT through the grants UID/FIS/04434/2013 and POCI-01-0145-FEDER-007672.




\bibliographystyle{mnras}
\bibliography{sousa_bibliography} 

\begin{thebibliography}{}
\makeatletter
\relax
\def\mn@urlcharsother{\let\do\@makeother \do\$\do\&\do\#\do\^\do\_\do\%\do\~}
\def\mn@doi{\begingroup\mn@urlcharsother \@ifnextchar [ {\mn@doi@}
  {\mn@doi@[]}}
\def\mn@doi@[#1]#2{\def\@tempa{#1}\ifx\@tempa\@empty \href
  {http://dx.doi.org/#2} {doi:#2}\else \href {http://dx.doi.org/#2} {#1}\fi
  \endgroup}
\def\mn@eprint#1#2{\mn@eprint@#1:#2::\@nil}
\def\mn@eprint@arXiv#1{\href {http://arxiv.org/abs/#1} {{\tt arXiv:#1}}}
\def\mn@eprint@dblp#1{\href {http://dblp.uni-trier.de/rec/bibtex/#1.xml}
  {dblp:#1}}
\def\mn@eprint@#1:#2:#3:#4\@nil{\def\@tempa {#1}\def\@tempb {#2}\def\@tempc
  {#3}\ifx \@tempc \@empty \let \@tempc \@tempb \let \@tempb \@tempa \fi \ifx
  \@tempb \@empty \def\@tempb {arXiv}\fi \@ifundefined
  {mn@eprint@\@tempb}{\@tempb:\@tempc}{\expandafter \expandafter \csname
  mn@eprint@\@tempb\endcsname \expandafter{\@tempc}}}

\bibitem[\protect\citeauthoryear{{Adibekyan}, {Sousa}, {Santos}, {Delgado
  Mena}, {Gonz{\'a}lez Hern{\'a}ndez}, {Israelian}, {Mayor}  \&
  {Khachatryan}}{{Adibekyan} et~al.}{2012}]{Adibekyan-2012b}
{Adibekyan} V.~Z.,  {Sousa} S.~G.,  {Santos} N.~C.,  {Delgado Mena} E.,
  {Gonz{\'a}lez Hern{\'a}ndez} J.~I.,  {Israelian} G.,  {Mayor} M.,
  {Khachatryan} G.,  2012, \mn@doi [\aap] {10.1051/0004-6361/201219401}, \href
  {http://adsabs.harvard.edu/abs/2012A%26A...545A..32A} {545, A32}

\bibitem[\protect\citeauthoryear{{Baruteau}, {Bai}, {Mordasini}  \&
  {Molli{\`e}re}}{{Baruteau} et~al.}{2016}]{Baruteau-2016}
{Baruteau} C.,  {Bai} X.,  {Mordasini} C.,   {Molli{\`e}re} P.,  2016, \mn@doi
  [\ssr] {10.1007/s11214-016-0258-z}, \href
  {http://adsabs.harvard.edu/abs/2016SSRv..205...77B} {205, 77}

\bibitem[\protect\citeauthoryear{{Becker} \& {Adams}}{{Becker} \&
  {Adams}}{2017}]{Becker-2017}
{Becker} J.~C.,  {Adams} F.~C.,  2017, \mn@doi [\mnras] {10.1093/mnras/stx461},
  \href {http://adsabs.harvard.edu/abs/2017MNRAS.468..549B} {468, 549}

\bibitem[\protect\citeauthoryear{{Bou{\'e}}, {Figueira}, {Correia}  \&
  {Santos}}{{Bou{\'e}} et~al.}{2012}]{Boue-2012}
{Bou{\'e}} G.,  {Figueira} P.,  {Correia} A.~C.~M.,   {Santos} N.~C.,  2012,
  \mn@doi [\aap] {10.1051/0004-6361/201118084}, \href
  {http://adsabs.harvard.edu/abs/2012A%26A...537L...3B} {537, L3}

\bibitem[\protect\citeauthoryear{{Buchhave} \& {Latham}}{{Buchhave} \&
  {Latham}}{2015}]{Buchhave-2015}
{Buchhave} L.~A.,  {Latham} D.~W.,  2015, \mn@doi [\apj]
  {10.1088/0004-637X/808/2/187}, \href
  {http://adsabs.harvard.edu/abs/2015ApJ...808..187B} {808, 187}

\bibitem[\protect\citeauthoryear{{Buchhave} et~al.,}{{Buchhave}
  et~al.}{2012}]{Buchhave-2012}
{Buchhave} L.~A.,  et~al., 2012, \mn@doi [\nat] {10.1038/nature11121}, \href
  {http://adsabs.harvard.edu/abs/2012Natur.486..375B} {486, 375}

\bibitem[\protect\citeauthoryear{{Buchhave} et~al.,}{{Buchhave}
  et~al.}{2014}]{Buchhave-2014}
{Buchhave} L.~A.,  et~al., 2014, \mn@doi [\nat] {10.1038/nature13254}, \href
  {http://adsabs.harvard.edu/abs/2014Natur.509..593B} {509, 593}

\bibitem[\protect\citeauthoryear{{Chiang} \& {Laughlin}}{{Chiang} \&
  {Laughlin}}{2013}]{Chiang-2013}
{Chiang} E.,  {Laughlin} G.,  2013, \mn@doi [\mnras] {10.1093/mnras/stt424},
  \href {http://adsabs.harvard.edu/abs/2013MNRAS.431.3444C} {431, 3444}

\bibitem[\protect\citeauthoryear{{Courcol}, {Bouchy}  \& {Deleuil}}{{Courcol}
  et~al.}{2016}]{Courcol-2016}
{Courcol} B.,  {Bouchy} F.,   {Deleuil} M.,  2016, \mn@doi [\mnras]
  {10.1093/mnras/stw1049}, \href
  {http://adsabs.harvard.edu/abs/2016MNRAS.461.1841C} {461, 1841}

\bibitem[\protect\citeauthoryear{{Dawson} \& {Johnson}}{{Dawson} \&
  {Johnson}}{2018}]{Dawson-2018}
{Dawson} R.~I.,  {Johnson} J.~A.,  2018, preprint, \href
  {http://adsabs.harvard.edu/abs/2018arXiv180106117D} {} (\mn@eprint {arXiv}
  {1801.06117})

\bibitem[\protect\citeauthoryear{{Fischer} \& {Valenti}}{{Fischer} \&
  {Valenti}}{2005}]{Fischer_Valenti-2005}
{Fischer} D.~A.,  {Valenti} J.,  2005, \mn@doi [\apj] {10.1086/428383}, \href
  {http://adsabs.harvard.edu/cgi-bin/nph-bib_query?bibcode=2005ApJ...622.1102F&db_key=AST}
  {622, 1102}

\bibitem[\protect\citeauthoryear{Foreman-Mackey}{Foreman-Mackey}{2016}]{corner}
Foreman-Mackey D.,  2016, \mn@doi [The Journal of Open Source Software]
  {10.21105/joss.00024}, 24

\bibitem[\protect\citeauthoryear{{Ghezzi}, {Cunha}, {Smith}, {de Ara{\'u}jo},
  {Schuler}  \& {de la Reza}}{{Ghezzi} et~al.}{2010}]{Ghezzi-2010}
{Ghezzi} L.,  {Cunha} K.,  {Smith} V.~V.,  {de Ara{\'u}jo} F.~X.,  {Schuler}
  S.~C.,   {de la Reza} R.,  2010, \mn@doi [\apj]
  {10.1088/0004-637X/720/2/1290}, \href
  {http://adsabs.harvard.edu/abs/2010ApJ...720.1290G} {720, 1290}

\bibitem[\protect\citeauthoryear{{Hogg}, {Bovy}  \& {Lang}}{{Hogg}
  et~al.}{2010}]{Hogg-2010}
{Hogg} D.~W.,  {Bovy} J.,   {Lang} D.,  2010, arXiv e-prints, \href
  {http://adsabs.harvard.edu/abs/2010arXiv1008.4686H} {}

\bibitem[\protect\citeauthoryear{{Huang}, {Wu}  \& {Triaud}}{{Huang}
  et~al.}{2016}]{Huang-2016}
{Huang} C.,  {Wu} Y.,   {Triaud} A.~H.~M.~J.,  2016, \mn@doi [\apj]
  {10.3847/0004-637X/825/2/98}, \href
  {http://adsabs.harvard.edu/abs/2016ApJ...825...98H} {825, 98}

\bibitem[\protect\citeauthoryear{{Kobayashi}, {Ormel}  \& {Ida}}{{Kobayashi}
  et~al.}{2012}]{Kobayashi-2012}
{Kobayashi} H.,  {Ormel} C.~W.,   {Ida} S.,  2012, \mn@doi [\apj]
  {10.1088/0004-637X/756/1/70}, \href
  {http://adsabs.harvard.edu/abs/2012ApJ...756...70K} {756, 70}

\bibitem[\protect\citeauthoryear{{Lubow} \& {Ida}}{{Lubow} \&
  {Ida}}{2010}]{Lubow-2010}
{Lubow} S.~H.,  {Ida} S.,  2010, preprint, \href
  {http://adsabs.harvard.edu/abs/2010arXiv1004.4137L} {} (\mn@eprint {arXiv}
  {1004.4137})

\bibitem[\protect\citeauthoryear{{Mann}, {Brewer}, {Gaidos}, {L{\'e}pine}  \&
  {Hilton}}{{Mann} et~al.}{2013}]{Mann-2013}
{Mann} A.~W.,  {Brewer} J.~M.,  {Gaidos} E.,  {L{\'e}pine} S.,   {Hilton}
  E.~J.,  2013, \mn@doi [\aj] {10.1088/0004-6256/145/2/52}, \href
  {http://adsabs.harvard.edu/abs/2013AJ....145...52M} {145, 52}

\bibitem[\protect\citeauthoryear{{Mayor} et~al.,}{{Mayor}
  et~al.}{2011}]{Mayor-2011}
{Mayor} M.,  et~al., 2011, preprint, \href
  {http://adsabs.harvard.edu/abs/2011arXiv1109.2497M} {} (\mn@eprint {arXiv}
  {1109.2497})

\bibitem[\protect\citeauthoryear{{Munoz Romero} \& {Kempton}}{{Munoz Romero} \&
  {Kempton}}{2018}]{Munoz-Romero-2018}
{Munoz Romero} C.~E.,  {Kempton} E.~M.-R.,  2018, \mn@doi [\aj]
  {10.3847/1538-3881/aaab5e}, \href
  {http://adsabs.harvard.edu/abs/2018AJ....155..134M} {155, 134}

\bibitem[\protect\citeauthoryear{{{\"O}nehag}, {Heiter}, {Gustafsson},
  {Piskunov}, {Plez}  \& {Reiners}}{{{\"O}nehag} et~al.}{2012}]{Onehag-2012}
{{\"O}nehag} A.,  {Heiter} U.,  {Gustafsson} B.,  {Piskunov} N.,  {Plez} B.,
  {Reiners} A.,  2012, \mn@doi [\aap] {10.1051/0004-6361/201118101}, \href
  {http://adsabs.harvard.edu/abs/2012A%26A...542A..33O} {542, A33}

\bibitem[\protect\citeauthoryear{{Petigura} et~al.,}{{Petigura}
  et~al.}{2018}]{Petigura-2018}
{Petigura} E.~A.,  et~al., 2018, \mn@doi [\aj] {10.3847/1538-3881/aaa54c},
  \href {http://adsabs.harvard.edu/abs/2018AJ....155...89P} {155, 89}

\bibitem[\protect\citeauthoryear{{Rojas-Ayala}, {Covey}, {Muirhead}  \&
  {Lloyd}}{{Rojas-Ayala} et~al.}{2012}]{Rojas-Ayala-2012}
{Rojas-Ayala} B.,  {Covey} K.~R.,  {Muirhead} P.~S.,   {Lloyd} J.~P.,  2012,
  \mn@doi [\apj] {10.1088/0004-637X/748/2/93}, \href
  {http://adsabs.harvard.edu/abs/2012ApJ...748...93R} {748, 93}

\bibitem[\protect\citeauthoryear{{Santerne} et~al.,}{{Santerne}
  et~al.}{2016}]{Santerne-2016b}
{Santerne} A.,  et~al., 2016, \mn@doi [\aap] {10.1051/0004-6361/201527329},
  \href {http://adsabs.harvard.edu/abs/2016A%26A...587A..64S} {587, A64}

\bibitem[\protect\citeauthoryear{{Santos}, {Israelian}  \& {Mayor}}{{Santos}
  et~al.}{2004}]{Santos-2004b}
{Santos} N.~C.,  {Israelian} G.,   {Mayor} M.,  2004, A\&A, \href
  {http://adsabs.harvard.edu/cgi-bin/nph-bib_query?bibcode=2004A%26A...415.1153S&amp;db_key=AST}
  {415, 1153}

\bibitem[\protect\citeauthoryear{{Santos} et~al.,}{{Santos}
  et~al.}{2013}]{Santos-2013}
{Santos} N.~C.,  et~al., 2013, \mn@doi [\aap] {10.1051/0004-6361/201321286},
  \href {http://adsabs.harvard.edu/abs/2013A%26A...556A.150S} {556, A150}

\bibitem[\protect\citeauthoryear{{Schlaufman} \& {Winn}}{{Schlaufman} \&
  {Winn}}{2016}]{Schlaufman-2016}
{Schlaufman} K.~C.,  {Winn} J.~N.,  2016, \mn@doi [\apj]
  {10.3847/0004-637X/825/1/62}, \href
  {http://adsabs.harvard.edu/abs/2016ApJ...825...62S} {825, 62}

\bibitem[\protect\citeauthoryear{{Schneider}, {Dedieu}, {Le Sidaner}, {Savalle}
   \& {Zolotukhin}}{{Schneider} et~al.}{2011}]{Schneider-2011}
{Schneider} J.,  {Dedieu} C.,  {Le Sidaner} P.,  {Savalle} R.,   {Zolotukhin}
  I.,  2011, \mn@doi [\aap] {10.1051/0004-6361/201116713}, \href
  {http://adsabs.harvard.edu/abs/2011A%26A...532A..79S} {532, A79}

\bibitem[\protect\citeauthoryear{{Sousa}}{{Sousa}}{2014}]{Sousa-2014}
{Sousa} S.~G.,  2014, ArXiv e-prints - http://arxiv.org/abs/1407.5817, \href
  {http://adsabs.harvard.edu/abs/2014arXiv1407.5817S} {}

\bibitem[\protect\citeauthoryear{{Sousa} et~al.,}{{Sousa}
  et~al.}{2008}]{Sousa-2008}
{Sousa} S.~G.,  et~al., 2008, \mn@doi [A\&A] {10.1051/0004-6361:200809698},
  \href {http://adsabs.harvard.edu/abs/2008A%26A...487..373S} {487, 373}

\bibitem[\protect\citeauthoryear{{Sousa}, {Santos}, {Israelian}, {Lovis},
  {Mayor}, {Silva}  \& {Udry}}{{Sousa} et~al.}{2011a}]{Sousa-2011}
{Sousa} S.~G.,  {Santos} N.~C.,  {Israelian} G.,  {Lovis} C.,  {Mayor} M.,
  {Silva} P.~B.,   {Udry} S.,  2011a, \mn@doi [\aap]
  {10.1051/0004-6361/201015646}, \href
  {http://adsabs.harvard.edu/abs/2011A%26A...526A..99S} {526, A99+}

\bibitem[\protect\citeauthoryear{{Sousa}, {Santos}, {Israelian}, {Mayor}  \&
  {Udry}}{{Sousa} et~al.}{2011b}]{Sousa-2011b}
{Sousa} S.~G.,  {Santos} N.~C.,  {Israelian} G.,  {Mayor} M.,   {Udry} S.,
  2011b, \mn@doi [\aap] {10.1051/0004-6361/201117699}, \href
  {http://adsabs.harvard.edu/abs/2011A%26A...533A.141S} {533, A141}

\bibitem[\protect\citeauthoryear{{Sousa} et~al.,}{{Sousa}
  et~al.}{2018}]{Sousa-2018}
{Sousa} S.~G.,  et~al., 2018, preprint, \href
  {http://adsabs.harvard.edu/abs/2018arXiv181008108S} {} (\mn@eprint {arXiv}
  {1810.08108})

\bibitem[\protect\citeauthoryear{{Steffen} et~al.,}{{Steffen}
  et~al.}{2012}]{Steffen-2012}
{Steffen} J.~H.,  et~al., 2012, \mn@doi [Proceedings of the National Academy of
  Science] {10.1073/pnas.1120970109}, \href
  {http://adsabs.harvard.edu/abs/2012PNAS..109.7982S} {109, 7982}

\bibitem[\protect\citeauthoryear{{Udry} \& {Santos}}{{Udry} \&
  {Santos}}{2007}]{Udry-2007b}
{Udry} S.,  {Santos} N.~C.,  2007, \mn@doi [\araa]
  {10.1146/annurev.astro.45.051806.110529}, \href
  {http://adsabs.harvard.edu/abs/2007ARA%26A..45..397U} {45, 397}

\bibitem[\protect\citeauthoryear{{Udry} et~al.,}{{Udry}
  et~al.}{2006}]{Udry-2006}
{Udry} S.,  et~al., 2006, \mn@doi [A\&A] {10.1051/0004-6361:20054084}, \href
  {http://adsabs.harvard.edu/cgi-bin/nph-bib_query?bibcode=2006A%26A...447..361U&db_key=AST}
  {447, 361}

\bibitem[\protect\citeauthoryear{{Valenti} \& {Fischer}}{{Valenti} \&
  {Fischer}}{2008}]{Valenti-2008}
{Valenti} J.,  {Fischer} D.,  2008, in {van Belle} G.,  ed.,  Astronomical
  Society of the Pacific Conference Series Vol. 384, 14th Cambridge Workshop on
  Cool Stars, Stellar Systems, and the Sun. p.~292

\bibitem[\protect\citeauthoryear{{Wang} \& {Fischer}}{{Wang} \&
  {Fischer}}{2015}]{Wang-2015}
{Wang} J.,  {Fischer} D.~A.,  2015, \mn@doi [\aj] {10.1088/0004-6256/149/1/14},
  \href {http://adsabs.harvard.edu/abs/2015AJ....149...14W} {149, 14}

\bibitem[\protect\citeauthoryear{{Zhu}, {Wang}  \& {Huang}}{{Zhu}
  et~al.}{2016}]{Zhu-2016}
{Zhu} W.,  {Wang} J.,   {Huang} C.,  2016, \mn@doi [\apj]
  {10.3847/0004-637X/832/2/196}, \href
  {http://adsabs.harvard.edu/abs/2016ApJ...832..196Z} {832, 196}

\makeatother
\end{thebibliography}




\appendix
\section{Fitting the MPM diagram}
%
%

The 3D plane fitting was done using an MCMC approach following very closely and adapting the Python implemention presented in https://dfm.io/posts/fitting-a-plane/ which is based on Chapter 7 of \citet[][]{Hogg-2010}. Using this method we are able to use errors for all the variables and provide reliable errors estimations for retrieved coefficients using the data presented in Figure \ref{the_figure_low_mass}.

The general plane equation used to fit the data on the MPM diagram is written as:

\begin{equation}
 M_{p} = b + m1 * [Fe/H] + m2 * \log(P)
\label{eq_mass_gen}
\end{equation}

where Mp is the mass of the planet, [Fe/H] is the metallicity (in dex), P the period of the planet (in days), m1 and m2 are the slopes of the correlations and b is the constant value at the origin.

\begin{figure*}
  \centering
  \includegraphics[width=16cm]{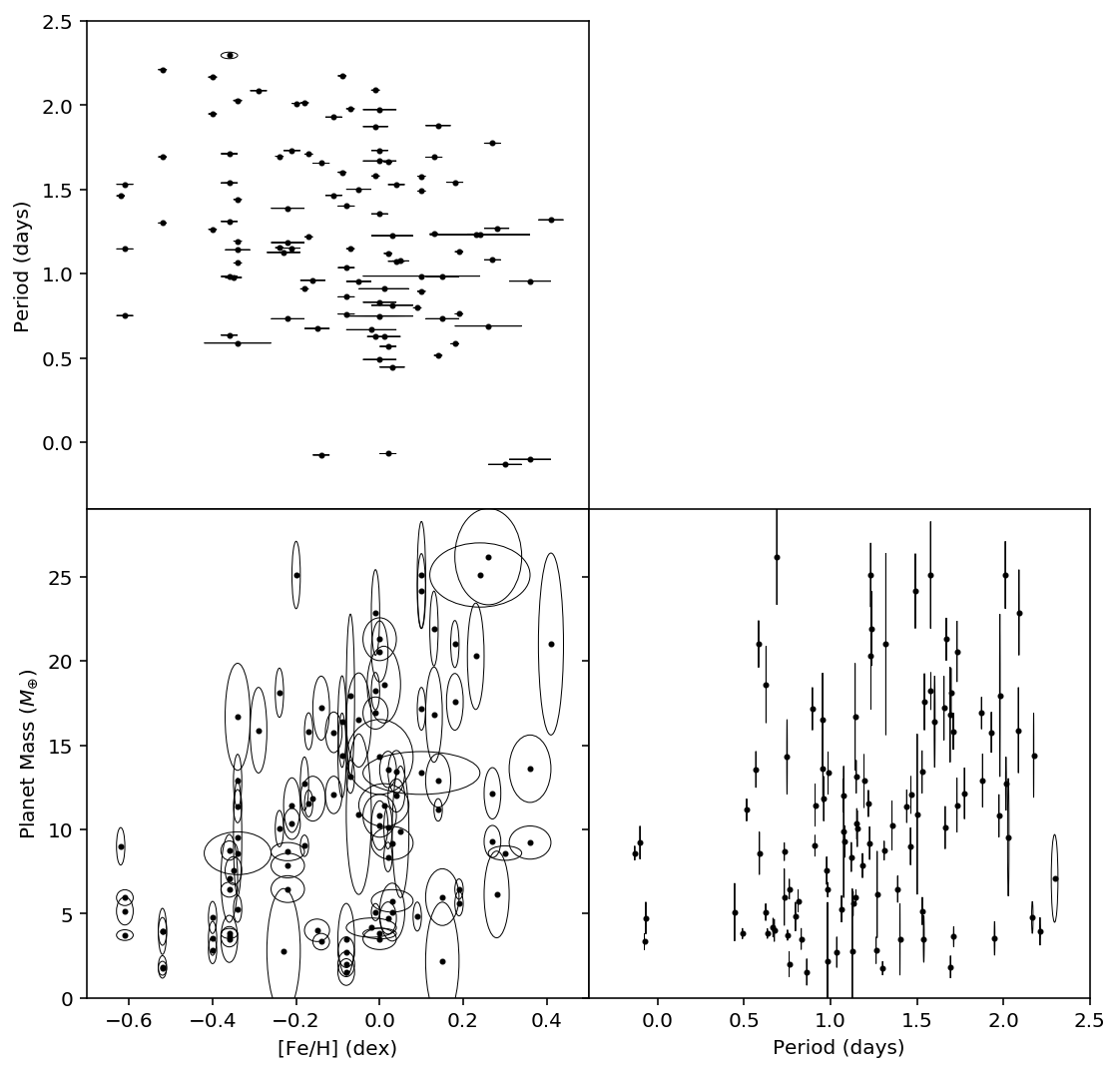}
  \caption{Data of Figure \ref{the_figure_low_mass} with respective errors.}
  \label{data_errors}
\end{figure*}

\begin{figure*}
  \centering
  \includegraphics[width=16cm]{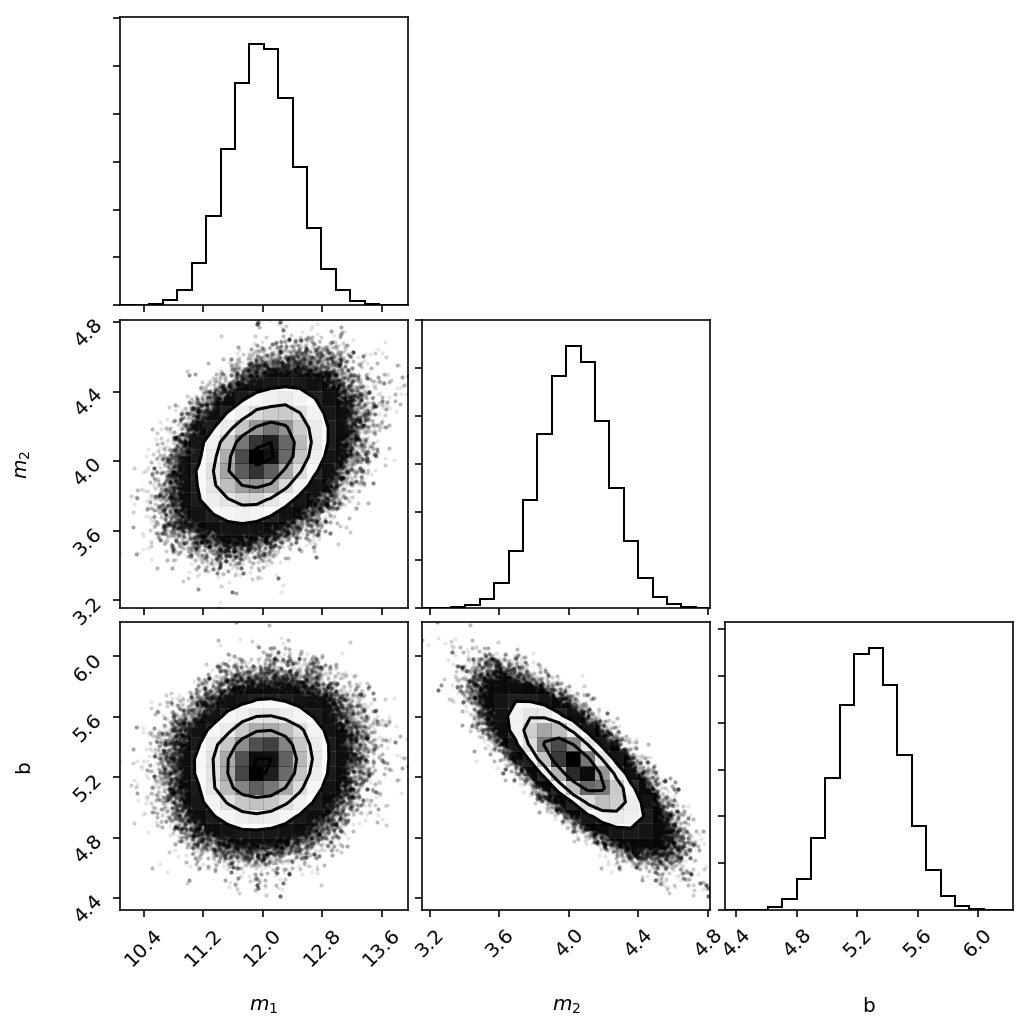}
  \caption{Samples to estimate the fitting coefficients of the plane correlation. This figure was produced using corner python module \citep[][]{corner}.}
  \label{data_samples}
\end{figure*}

\section{Simulated MPM diagram example}

\begin{figure}
  \centering
  \includegraphics[width=9cm]{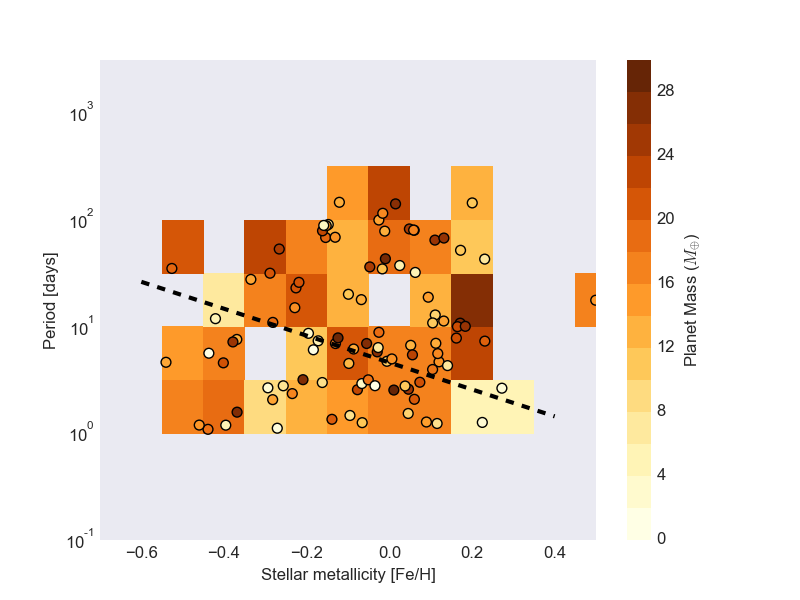} 
  \includegraphics[width=9cm]{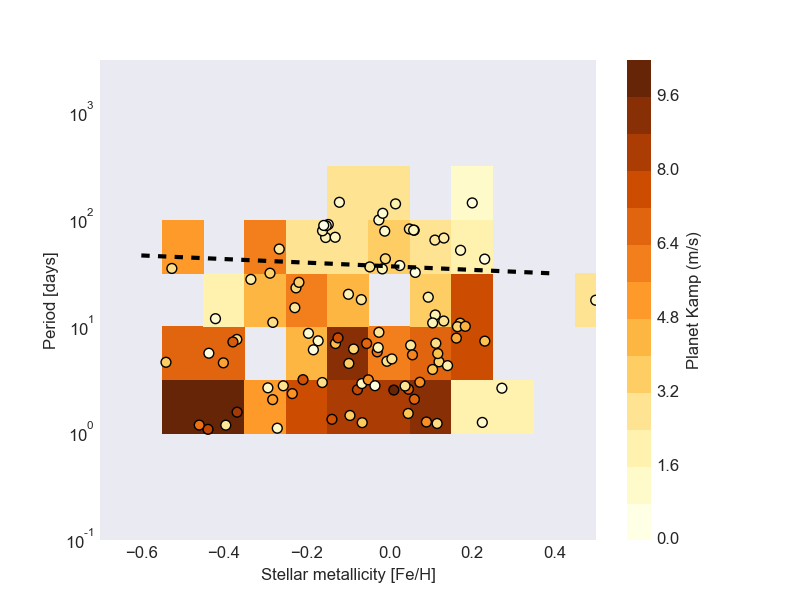}
  \caption{Same as Figure \ref{the_figure_low_mass} but for one simulated data (top panel) and the RV semi-amplitude (bottom panel). The dashed line represents the linear fit for a constant value, 10 M$_\oplus$ and 4 m/s on the top and bottom panels respectively.}
  \label{fig_simulated}
\end{figure}

\bsp	
\label{lastpage}
\end{document}